# Experimental and Computational Studies on the Reactivity of Methanimine Radical Cation (H$_2$CNH$^{+•}$) and its Isomer Aminomethylene (HCNH$_2$$^{+•}$) With C$_2$H$_2$

*Vincent Richardson[1], Daniela Ascenzi[1]\*, David Sundelin[2], Christian Alcaraz[3,4], Claire Romanzin[3,4], Roland Thissen[3,4], Jean-Claude Guillemin[5], Miroslav Polášek[6], Paolo Tosi[1], Jan Žabka[6] and Wolf D. Geppert[2]*

[1]Department of Physics, University of Trento, Trento, Italy, [2]Department of Physics, Stockholm University, Stockholm, Sweden, [3]CNRS, Institut de Chimie Physique, Université Paris-Saclay, Saint Aubin, France, [4]Synchrotron Soleil, L'Orme des Merisiers, Saint Aubin, France, [5]CNRS, Ecole Nationale Supérieure de Chimie de Rennes, Univ. Rennes, Rennes, France, [6]J. Heyrovsky Institute of Physical Chemistry of the Czech Academy of Sciences, Prague, Czechia



Experimental and theoretical studies are presented on the reactivity of the radical cation isomers H$_2$CNH$^{+•}$ (methanimine) and HCNH$_2$$^{+•}$ (aminomethylene) with ethyne (C$_2$H$_2$). Selective isomer generation is performed via dissociative photoionization of suitable neutral precursors as well as via direct photoionization of methanimine. Reactive cross sections (in absolute scales) and product branching ratios are measured as a function of photon and collision energies. Differences between isomers' reactivity are discussed in light of *ab-initio* calculations of reaction mechanisms. The major channels, for both isomers, are due to H atom elimination from covalently bound adducts to give [C$_3$NH$_4$]$^+$. Theoretical calculations show that while for the reaction of HCNH$_2$$^{+•}$ with acetylene any of the three lowest energy [C$_3$NH$_4$]$^+$ isomers can form via barrierless and exothermic pathways, for the H$_2$CNH$^{+•}$ reagent the only barrierless pathway is the one leading to the production of protonated vinyl cyanide (CH$_2$CHCNH$^+$), a prototypical branched nitrile species that has been proposed as a likely intermediate in star forming regions and in the atmosphere of Titan. The astrochemical implications of the results are briefly addressed.

Keywords: methylene imine, acrylonitrile, vinyl cyanide, rate constant, cross sections, Titan, ISM, ion-molecule reactions

## 1 INTRODUCTION

The reaction of ions is thought to play a crucial role in the synthesis of complex organic species in both the interstellar medium (ISM) and the atmospheres of planets and their satellites Larsson et al. (2012). As increasingly complex species are identified in such environments, the potential impact of isomers increases as well. The existence of isomeric ions in the ISM has been known for a long time and in some cases, such as HCO$^+$ and HOC$^+$, a determination has been made of their relative abundances Woods et al. (1983). As isomers have different spectroscopic and chemical properties, and, in many cases, isomerization barriers are too high to be overcome in interstellar conditions, they must be treated separately.





Notably, the relative abundance of isomers in a given environment cannot usually be determined by their thermodynamic properties, with the feasibility of formation and destruction mechanisms often more important than relative energies. Furthermore, while isomers often have quite different dipole moments and other spectroscopic properties and therefore have distinct radioastronomic profiles Woods et al. (1983), *in situ* measurements of ion abundances in planetary and satellite atmospheres often employ mass spectrometers which bring extreme sensitivity and time/localisation specific information, but cannot distinguish between isomers. The novel Atacama Large Interferometer Array (ALMA) has enabled data collection on the density as a function of altitude for HCN and HNC Cordiner et al. (2019) and the unprecedented resolution of this device should also allow for the retrieval of data about the abundance and distribution of isomeric ions in different astronomic objects. However, at present, the relative abundance of isomers in such environments has proven difficult to assess. So, in order to develop accurate models to predict these abundances, it is important to have reliable data on the reactivity of ionic isomers with common interstellar and atmospheric molecules.

In recent years, efforts have been made to selectively generate relevant isomers Fathi et al. (2016), with VUV photoionization in particular proving an effective tool Polášek et al. (2016) while noble gas tagging has been successful in characterizing the presence and the relative abundance of ions produced by electron impact ionization Brünken et al. (2019).

Like Earth, Titan has a dense nitrogen-dominated atmosphere Hörst (2017), which is one of the most chemically complex atmospheres in the solar system, as demonstrated by the data from the ion and neutral mass-spectrometer (INMS) of the Cassini mission. Titan's atmospheric chemistry is driven by the ionization and/or dissociation of its two primary components, $N_2$ and $CH_4$, through a combination of extreme ultraviolet (EUV) radiation Imanaka and Smith (2007) and magnetospheric electrons. The resultant ions and radicals can then react to form more complex species Vigren et al. (2012), Westlake et al. (2014), including large hydrocarbons and nitrogen-bearing compounds Vinatier et al. (2007), Nixon et al. (2018), with cations of masses up to $m/z$ 99 having been identified at altitudes of 950 km above the surface while cations of masses up to $m/z$ 350 were detected by the Cassini plasma spectrometer ion beam sensor (CAPS-IBS) Vuitton et al. (2007).

Reactions of nitrogen-containing ions with hydrocarbons, especially unsaturated hydrocarbons such as $C_2H_2$ and $C_2H_4$, have been suggested as the basis for the formation of some of the complex ions that have been detected in Titan's atmosphere Westlake et al. (2014), Vuitton et al. (2019). However, protonated nitriles, a major part of the nitrogen-containing ions in Titan's atmosphere, are less reactive with their major reaction pathway in Titan's atmosphere being dissociative recombination Vuitton et al. (2007). However, this is not the case for the methylenimine cation $H_2CNH^{+\bullet}$ and its isomer the aminomethylene ion $HCNH_2^{+\bullet}$, which are both reactive radical cations and could form the basis of chain elongation reactions with unsaturated and saturated hydrocarbons that produce larger ions, from which neutral species can then be generated by dissociative recombination. Though the $m/z$ 29 signal recorded by INMS is regarded as mostly arising from the presence of $C_2H_5^+$, models predict a density of $HCNH_2^{+\bullet}$ and its isomers amounting to $1.1 \times 10^{-2}$ cm$^{-3}$ in Titan's ionosphere Vuitton et al. (2007).

Although the $H_2CNH^{+\bullet}$ and $HCNH_2^{+\bullet}$ ions have not yet been identified in the ISM, neutral methanimine (*i.e.* methylene imine, $H_2CNH$) is ubiquitous, having been detected in several objects including giant molecular clouds as well as both high mass and solar-type protostellar systems Dickens et al. (1997), Suzuki et al. (2016), Widicus Weaver et al. (2017), Ligterink et al. (2018), Bogelund et al. (2019). Methanimine, which could form either on the surface of dust grains Bernstein et al. (1995), Theule et al. (2011) or via gas phase reactions Suzuki et al. (2016), could serve as the basis for synthetic routes to form more complex nitrogen-containing molecules, including biomolecule precursors such as glycine and its corresponding α-aminonitrile Basiuk and Bogillo (2002), Aponte et al. (2017).

Due to its low ionization energy of 9.99 eV Holzmeier et al. (2013), methanimine can be ionized by VUV photons in the ISM. A possible pathway for the synthesis of larger nitrogen-containing species could therefore be the ionization of $H_2CNH$ followed by the reaction of the resulting radical cation with an unsaturated hydrocarbon molecule, such as ethene or ethyne. The resultant ion could then either undergo dissociative recombination to form a complex neutral molecule through elimination of a hydrogen atom, or react further with other interstellar compounds to generate an even larger ionic species. Since methanimine has been detected also in ion-rich regions [e.g., at the "radical-ion peak" in the Orion molecular cloud Dickens et al. (1997)], such a course of reaction does not seem unlikely.

This paper presents a reactivity study of both $H_2CNH^{+\bullet}$ and $HCNH_2^{+\bullet}$ with ethyne using synchrotron radiation and apt neutral precursors to selectively generate the charged species. It aims to establish the occurrence of product channels leading to the formation of new C-C or C-N bonds through chain elongation reactions with ethyne, thereby leading to an increase in chemical complexity.

In **Section 2** and **Section 3** we present the experimental and theoretical methodologies. In **Section 4** we provide a detailed description of the experimental results while in **Section 5** the computational results are presented alongside proposed reaction mechanisms. Experimental results are discussed and interpreted in light of calculations of the most probable reaction mechanisms in **Section 6**. Finally, the conclusions are summarized in **Section 7**, where implications of present results on the formation of complex N-containing ions in Titan's atmosphere and in the ISM are briefly addressed.

## 2 EXPERIMENTAL METHODOLOGY

Experiments were performed using the CERISES apparatus, an associated experiment of the DESIRS beamline Nahon et al. (2012) of the SOLEIL synchrotron radiation facility. Since the experimental set-up has been described in detail previously





Alcaraz et al. (2004), Cunha de Miranda et al. (2015), only the most relevant details will be given here. CERISES is a guided ion beam tandem mass spectrometer composed of two octopoles located between two quadrupole mass filters in a Q1-O1-O2-Q2 configuration that permits investigation of bimolecular reactions of mass-selected ions occurring in a 4 cm long reaction cell located after the first octopole. Absolute reaction cross sections (CSs) and branching ratios (BRs) as a function of both photon ($E_{phot}$) and collision ($E_{CM}$) energies are derived by measuring the yields of parent- and product-ions.

The HCNH$_2^{+\bullet}$ and H$_2$CNH$^{+\bullet}$ isomers were produced by dissociative photoionization of the gaseous precursors cyclopropylamine (c-C$_3$H$_5$NH$_2$) and azetidine (c-CH$_2$CH$_2$CH$_2$NH) respectively, according to the following reactions:

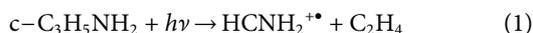
$$c-C_3H_5NH_2 + h\nu \rightarrow HCNH_2^{+\bullet} + C_2H_4 \qquad (1)$$

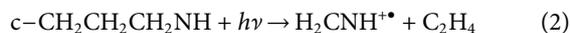
$$c-CH_2CH_2CH_2NH + h\nu \rightarrow H_2CNH^{+\bullet} + C_2H_4 \qquad (2)$$

These commercially available gaseous precursors were introduced into the ion source at roughly $10^{-6}$ mbar [see Richardson et al. (2021), Sundelin et al. (2021) for a complete description of the generation of the ions]. In addition, photoionization of methanimine (H$_2$CNH) was used to selectively generate the H$_2$CNH$^{+\bullet}$ isomer, with the details of the ion generation processes presented in **Section 4**.

The vacuum ultraviolet (VUV) radiation for photoionization is provided by the undulator-based DESIRS beamline Nahon et al. (2012), operating between 5 and 40 eV. Photons at the desired energies are selected and scanned simultaneously with the undulator peak energy by a normal incidence monochromator equipped with a low dispersion uncoated SiC grating (200 grooves/mm) optimized to provide photon flux up to $10^{13}$ photons/s. In the present experiments, the photon energies required to produce the [CH$_3$N]$^{+\bullet}$ isomer ions are in the region 9.5–14.0 eV. The monochromator slits were set in the range 300–600 $\mu$m, corresponding to a resolution of 20–40 meV, in order to balance the maximization of signal with avoiding detector saturation. The monochromator was operated at its first order of diffraction. Higher orders were removed from the incident beam using a gas filter Mercier et al. (2000) installed on the beam line and filled with Argon (at a pressure of 0.2 mbar) to efficiently remove any photons with $E_{phot}$ larger than 15.7 eV. The absolute scale of the photon energy was checked using sharp absorption lines of atomic Argon around 11.828 and 14.304 eV Minnhagen (1973), Kramida et al. (2020) that were observed with systematic shifts of about 10–20 meV above their tabulated values.

The resolution of the mass filters Q1 and Q2 were adjusted in order to provide a suitable compromise between signal and separation of adjacent masses. We estimate that in this way we are able to remove 99.99% of neighboring ions for Q1 and 99.90% of neighboring ions for Q2 at the lowest collision energy, with a loss of performance reaching 98.80% of neighboring ions at the highest acceleration potentials of 20 V.

The collision energy in the lab is dependent on the reagent ion charge (+1 in our case) and on the difference between the ion source and the reaction cell potentials. The retarding potential method was applied to the reagent ion beam Teloy and Gerlich (1974) to determine the maximum of the first derivative of the reagent ion yield, which defines the zero of the kinetic energy. In this way we have estimated an average reagent ion beam FWHM of ~0.4 eV, equivalent to ~0.19 eV in the center-of-mass frame. By changing the potentials of the reaction cell and all subsequent elements, we were able to scan a collision energy range from ~0.06 to ~10 eV in the center of mass frame ($E_{CM}$).

C$_2$H$_2$ was stored in a dilute mixture with acetone, which was removed using a coil cooled with dry ice before insertion into the instrument. C$_2$H$_2$ was introduced in the reaction cell at a dynamic pressure of up to $2.0 \times 10^{-7}$ bar, which guarantees operation close to a single collision regime, keeps secondary reactions at a reduced level and limits the attenuation of the reagent ion to less than 10%, while still providing an accurate pressure measurement and a dynamic range on the absolute cross sections determination in the range 0.01–100 Å$^2$. The absolute pressure in the reaction cell was measured using an MKS 398H differential manometer instrument. All data were collected with either reaction gas in the cell or in the surrounding chamber, therefore allowing the removal of any contribution resulting from collisions occurring outside the reaction cell, corresponding to no more than 5% of product ion counts. The absolute cross section acquisitions were performed in the so-called multi-scan mode, which involves recording signals for all ionic species at a single point before moving to the next point. This method has the benefit of drastically reducing the effects of potential drifts of source pressure, reaction cell pressure and photon flux.

## 3 THEORETICAL METHODOLOGY

The mechanisms for the reactions of H$_2$CNH$^{+\bullet}$ and HCNH$_2^{+\bullet}$ were studied using GAUSSIAN 09, Revision D.01 Frish et al. (2013). Geometries for intermediate structure, *i.e.* minima and transition states (TSs), were calculated at the MP2/6-31G(d) and MP2/6-311++G(d,p) levels. The identity of TS and minima were checked by frequency calculations and zero-point energy corrections were applied to the obtained energies. IRC calculations were performed at the MP2/6-31G(d) level to ensure that the TS connect the correct minima. Single point energy calculations were carried out for all stationary points at the CCSD(T)/6-311++G(d,p) level with zero-point energy corrections taken from the MP2/6-311++G(d,p) level. Reactants and product geometries and energies were also calculated at the same level (including zero-point energy corrections) and the reactants' energies subtracted from the optimizations of the TS and minima, resulting in the relative energies $E_{rel}$. Cartesian coordinates, structures and electronic energies of reagents, products, minima and TS are reported in the **Supplementary Material**.

## 4 EXPERIMENTAL RESULTS

The characterization of the fragmentation pathways for generation of the two isomers via dissociative photoionization





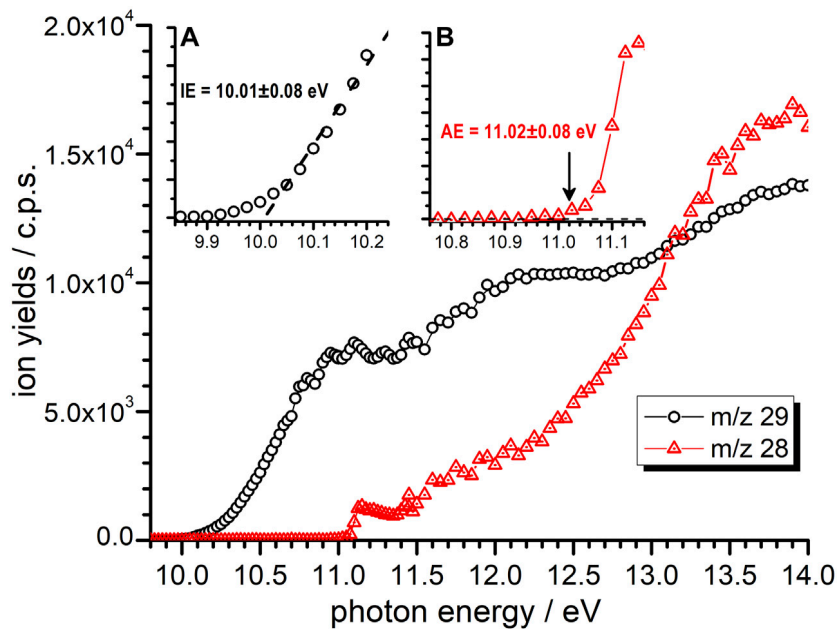

**FIGURE 1** | PIE curves for the photoionization of methanimine in the 9.8–14 eV photon energy region: open black circles are for $H_2CNH^{+\bullet}$ radical cations (*m/z* 29) and open red triangles for $HCNH^+$ dissociation products (*m/z* 28). Insets **(A)** and **(B)** show enlarged plots of the threshold regions for the corresponding *m/z* values. The dashed line indicates the linear extrapolation used to extract the experimental IE.

of cyclopropylamine (c-$C_3H_5NH_2$) and azetidine (c-$CH_2CH_2CH_2NH$) has been described in previous publications Richardson et al. (2021), Sundelin et al. (2021) to which the interested reader is referred to for details. In the following subsection we will focus on the formation of the $H_2CNH^{+\bullet}$ isomer through direct ionization of methanimine.

## 4.1 Generation of $H_2CNH^{+\bullet}$ From Photoionization of Methanimine

$H_2CNH$ was synthesized by dehydrocyanation of aminoacetonitrile performed under vacuum, following an improved version of the original method Guillemin and Denis (1988) and the interested reader is referred to Gans et al. (2019) for details of the chemical procedure. After synthesis, methanimine was stored under liquid nitrogen, and transferred to dry ice for introduction in the ion source of the CERISES set-up using its vapor pressure at dry ice temperature (T ~ −78°C).

The photoionization efficiency (PIE) curves (i.e. ion yields as a function of photon energy) for the parent radical cation $H_2CNH^{+\bullet}$ (*m/z* 29) and the photo-dissociation product $HCNH^+$ (*m/z* 28) are reported in **Figure 1**, where the observed Appearance Energies (AEs) are given. For $H_2CNH^{+\bullet}$, an Ionization Energy (IE) of 10.01 ± 0.08 eV was obtained using the method of linear threshold extrapolation of the PIE curve at *m/z* 29 Chupka (1971), Traeger and McLoughlin (1981), Castrovilli et al. (2014), a value that is fully consistent with the adiabatic ionization energy of methanimine of 9.99 eV as determined via mass-selected threshold photoelectron spectroscopy Holzmeier et al. (2013). For the $HCNH^+$ photo-fragment, an experimental AE = 11.02 ± 0.08 eV was estimated taking the first point above noise level of the PIE curve at *m/z* 28, after signal correction due to "spillover" of the more abundant peak at *m/z* 29. This is also consistent with the onset of dissociative photoionization of methanimine via $H^\bullet$ loss, observed at 11.13 eV Holzmeier et al. (2013).

We emphasize that the alternative generation method starting from neutral methanimine is explored as a way to produce the methanimine radical cation isomer free from interference from isobaric contaminants ($C_2H_5^+$, $H^{13}CNH^+$ and $^{13}CCH_4^{+\bullet}$), as is detailed in **Section 4.3**. Conversely, since the methanimine sample is free from heavier mass contaminants, the only expected interference at *m/z* 29 is limited to the $^{13}C$ contribution of the $HCNH^+$ photo-fragment at *m/z* 28. However, from **Figure 1** it can be clearly seen that this fragment is less intense than the *m/z* 29 cation at least up to a photon energy of ~13.1 eV. As a consequence, any contamination due to $H^{13}CNH^+$ at *m/z* 29 should account for less than 1% of the *m/z* 29 signal below $E_{phot}$ = 13.1 eV.

## 4.2 Reactions of $H_2CNH^{+\bullet}$ and $HCNH_2^{+\bullet}$ With $C_2H_2$

The reaction of both isomers yields products at *m/z* 54, 39 and 28, while for the $H_2CNH^{+\bullet}$ isomer an additional minor product was observed at *m/z* 27. The most probable product channels are indicated as **Eqs 3–11**, and the reaction enthalpies estimated from





**TABLE 1** | Reaction enthalpies for the reaction: HCNH$_2^{+•}$/H$_2$CNH$^{+•}$ + C$_2$H$_2$ → products.

| Products | React. No. | ΔH° with HCNH$_2^{+•}$ (eV)[a] | | ΔH° with H$_2$CNH$^{+•}$ (eV)[a] | |
|---|---|---|---|---|---|
| | | Lit. values (298 K)[a] | Calcs. this work (0 K) | Lit. values (298 K)[a] | Calcs. this work (0 K) |
| C$_2$H$_3^+$ + H$_2$CN$^•$ | Eq. 3 | 1.01 ± 0.15[b] | 1.08 | 0.85 ± 0.15[b] | 0.88 |
| C$_2$H$_3^+$ + HCNH$^•$ | Eq. 4 | 1.36 ± 0.15[b] | 1.50 | 1.20 ± 0.15[b] | 1.31 |
| HCNH$^+$ + C$_2$H$_3^•$ | Eq. 5 | −0.06 ± 0.31[c] | −0.09 | −0.23 ± 0.31[c] | −0.29 |
| C$_2$H$_4^{+•}$ + HCN | Eq. 6 | −0.57 ± 0.20[d] | −0.74 | −0.73 ± 0.20[d] | −0.93 |
| c-C$_3$H$_3^+$ + NH$_2^•$ | Eq. 7 | 0.05 ± 0.27[e] | 0.04 | −0.12 ± 0.27[e] | −0.16 |
| H$_2$CCHCNH$^+$ + H$^•$ | Eq. 8 | −1.18 ± 0.20[f] | −1.11 | −1.34 ± 0.20[f] | −1.31 |
| c-CHCHC(NH$_2$)$^+$ + H$^•$ | Eq. 9 | −0.96 ± 0.20[g] | −0.76 | −1.13 ± 0.20[g] | −0.96 |
| HCCCHNH$_2^+$ + H$^•$ | Eq. 10 | −0.79 ± 0.20[h] | −0.68 | −0.95 ± 0.20[h] | −0.87 |
| H$_2$CCHNCH$^+$ + H$^•$ | Eq. 11 | −0.67 ± 0.20[i] | −0.68 | −0.84 ± 0.20[i] | −0.87 |
| H$_2$CCCNH$_2^+$ + H$^•$ | Eq. 12 | −0.61 ± 0.20[j] | −0.24 | −0.78 ± 0.20[j] | −0.44 |
| CH$_2$CNCH$_2^+$ + H$^•$ | Eq. 13 | −0.54 ± 0.20[k] | −0.25 | −0.71 ± 0.20[k] | −0.45 |

[a]All values in the Table have been evaluated using Δ$_f$H°$_{298}$ (HCNH$_2^{+•}$) = 10.67 ± 0.1 eV Nguyen et al. (1994), Δ$_f$H°$_{298}$ (H$_2$CNH$^{+•}$) = 10.84 ± 0.1 eV Nguyen et al. (1994) and Δ$_f$H°$_{298}$ (C$_2$H$_2$) = 2.350 ± 0.009 Kramida et al. (2020).
[b]Calculated using the theoretical Δ$_f$H°$_{298}$ = 11.57 ± 0.03 eV for the vinyl cation CH$_2$CH$^+$ Lago and Baer (2006), Δ$_f$H°$_{298}$ (H$_2$CN$^•$) = 2.469 ± 0.007 eV Ruscic and Bross (2020) and Δ$_f$H°$_{298}$ (HCNH$^•$) = 2.821 ± 0.007 eV Ruscic and Bross (2020).
[c]Calculated assuming the formation of the vinyl radical (C$_2$H$_3^•$) with Δ$_f$H°$_{298}$ = 3.10 ± 0.05 eV Kramida et al. (2020) and HCNH$^+$ with Δ$_f$H°$_{298}$ = 9.87 ± 0.15 eV Holmes et al. (2006).
[d]Calculated using Δ$_f$H°$_{298}$ = 11.057 ± 0.006 eV for C$_2$H$_4^{+•}$ Kramida et al. (2020) and Δ$_f$H°$_{298}$ = 1.40 ± 0.10 eV for HCN Kramida et al. (2020).
[e]Assuming the formation of the c-C$_3$H$_3^+$ product with Δ$_f$H°$_{298}$ = 11.1 ± 0.10 eV Holmes et al. (2006) and the amino radical (NH$_2^•$) with Δ$_f$H°$_{298}$ = 1.97 ± 0.07 eV Kramida et al. (2020). The formation of the propargyl cation CH$_2$CCH$^+$ is not considered since this isomer lies ~1.1 eV higher in energy than the cyclic isomer, thus making the corresponding reaction endothermic.
[f]Assuming the formation of protonated vinyl cyanide H$_2$CCHCNH$^+$, with Δ$_f$H° = 9.59 ± 0.08 eV Holmes et al. (2006); Kramida et al. (2020).
[g]Assuming the formation of c-CHCHC(NH$_2$)$^+$, using a relative energy (with respect to the lowest energy isomer H$_2$CCHCNH$^+$) of 0.23 eV from calculations in Wang et al. (2015).
[h]Assuming the formation of HCCCHNH$_2^+$, using a relative energies (with respect to H$_2$CCHCNH$^+$) of 0.39 eV from calculations in Wang et al. (2015).
[i]Assuming the formation of H$_2$CCHNCH$^+$ (protonated isocyanoethene), using a relative energy (with respect to H$_2$CCHCNH$^+$) of 0.51 eV from calculations in Wang et al. (2015), Salpin et al. (1999).
[j]Assuming the formation of H$_2$CCCNH$_2^+$, using a relative energy (with respect to H$_2$CCHCNH$^+$) of 0.59 eV from calculations in Wang et al. (2015).
[k]Assuming the formation of CH$_2$CNCH$_2^+$, using a relative energy (with respect to H$_2$CCHCNH$^+$) of 0.633 eV from calculations in Wang et al. (2015).

literature values are summarized and compared with results from our theoretical calculations in **Table 1**.

$$H_2CNH^{+•}/HCNH_2^{+•} + C_2H_2 \rightarrow C_2H_3^+ \, (m/z \, 27) + H_2CN^• \quad (3)$$
$$\rightarrow C_2H_3^+ \, (m/z \, 27) + HCNH^• \quad (4)$$
$$\rightarrow HCNH^+ \, (m/z \, 28) + C_2H_3^• \quad (5)$$
$$\rightarrow C_2H_4^{+•} \, (m/z \, 28) + HCN \quad (6)$$
$$\rightarrow c-C_3H_3^+ \, (m/z \, 39) + NH_2^• \quad (7)$$
$$\rightarrow H_2CCHCNH^+ \, (m/z \, 54) + H^• \quad (8)$$
$$\rightarrow c-CHCHC(NH_2)^+ \, (m/z \, 54) + H^• \quad (9)$$
$$\rightarrow HCCCHNH_2^+ \, (m/z \, 54) + H^• \quad (10)$$
$$\rightarrow H_2CCHNCH^+ \, (m/z \, 54) + H^• \quad (11)$$
$$\rightarrow H_2CCCNH_2^+ \, (m/z \, 54) + H^• \quad (12)$$
$$\rightarrow CH_2CNCH_2^+ \, (m/z \, 54) + H^• \quad (13)$$

Although the *m/z* 39 product could alternatively correspond to the formation of HCCN$^{+•}$ or HCNC$^{+•}$ ions formed together with a CH$_4$ molecule, these processes are not only endothermic by about 2 eV [on the basis of the calculated energies of HCCN$^{+•}$ and HCNC$^{+•}$ Harland and McIntosh (1985), Goldberg et al. (1995), Holmes et al. (2006)] but also very difficult to account for from a mechanistic point of view. Hence this possibility is not considered any further and the corresponding products are not included in **Table 1**.

## 4.3 Considerations About Impurities in the Reagent Ions

When selecting *m/z* 29 from the neutral precursors (cyclopropylamine and azetidine), there is no way to discriminate against the presence of isobaric contaminants, namely C$_2$H$_5^+$, H$^{13}$CNH$^+$ and $^{13}$CCH$_4^{+•}$. Calculations and experimental evidence from our previous works Richardson et al. (2021), Sundelin et al. (2021) show that dissociative photoionization of azetidine to give H$_2$CNH$^{+•}$ plus C$_2$H$_4$ has an experimental AE (10.2 ± 0.1 eV) very close to the fragmentation channel giving HCNH$^+$ plus C$_2$H$_5^{+•}$. Additionally, the azetidine radical cation can fragment into C$_2$H$_4^{+•}$ plus H$_2$CNH at photon energies above ~11.3 eV. Hence both H$^{13}$CNH$^+$ and $^{13}$CCH$_4^{+•}$ might be present as impurities when mass selecting reagent ions at *m/z* 29 from azetidine. For cyclopropylamine, dissociative photoionization to HCNH$_2^{+•}$ (plus C$_2$H$_4$) has a similar AE (10.2 ± 0.1 eV) also very close to fragmentation into HCNH$^+$ plus C$_2$H$_5^•$. Additionally, the formation of the isobaric cation C$_2$H$_5^+$ is possible at photon energies higher than ~12.4 eV. Since fragmentation into C$_2$H$_4^{+•}$ plus HCNH$_2$ opens up at very high photon energies, any contamination due to $^{13}$CCH$_4^{+•}$ is not relevant in this case. Hence, for cyclopropylamine the relevant contaminants at *m/z* 29 are H$^{13}$CNH$^+$ and C$_2$H$_5^+$.

In the following the reactivity of such isobaric impurities with C$_2$H$_2$ will be addressed.





### 4.3.1 Contamination due to C$_2$H$_5^+$

Literature data Kim et al. (1977), Anicich (1993), Anicich (2003) indicate that C$_2$H$_5^+$ reacts with C$_2$H$_2$ with an overall rate $k = 1.9 \times 10^{-10}$ cm$^3 \cdot$molec$^{-1} \cdot$s$^{-1}$ to give the following products (with BRs indicated in parenthesis):

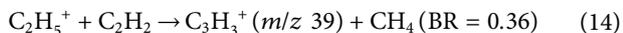
(14)

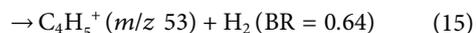
(15)

A recent re-evaluation Richardson et al. (n.d.) of this reaction using the same set-up employed in this work where C$_2$H$_5^+$ was generated via VUV dissociative photoionization of C$_2$H$_5$Br at a photon energy of 11.3 eV and at the lowest achievable collision energy (E$_{CM}$ ~ 0.08 eV) gives quite different product branching ratios. Most notably, **Eq. 14** is found to be the major channel (BR = 0.71), while **Eq. 15** is the second most abundant channel (BR = 0.22), with minor amounts of the other products (C$_4$H$_7^+$ and C$_2$H$_3^+$) accounting for the remaining product flux.

Since **Eq. 14** from C$_2$H$_5^+$ leads to a product of the same mass as **Eq. 7**, contamination at m/z 39 will emerge as the photon energy is increased when working with HCNH$_2^{+\bullet}$ from cyclopropylamine.

### 4.3.2 Contamination due to H$^{13}$CNH$^+$ and $^{13}$CCH$_4^{+\bullet}$

As already mentioned, a contribution from H$^{13}$CNH$^+$ at all photon energies is expected for both azetidine and cyclopropylamine experiments. However, literature reports indicate that HCNH$^+$ is unreactive with C$_2$H$_2$ [$k < 1.0 \times 10^{-13}$ cm$^3 \cdot$molec$^{-1} \cdot$s$^{-1}$ at room temperature Anicich et al. (2000), Anicich and McEwan (2001), Milligan et al. (2001)], as expected from thermochemical considerations, since the most feasible channel (i.e. proton transfer to give C$_2$H$_3^+$ plus HCN) is endothermic by approximately 0.75 eV. In our experiment, this channel could potentially open but only at high photon energies and should be accounted for when looking at data collected at m/z 27 under such conditions. Contamination due to $^{13}$CCH$_4^{+\bullet}$ is expected only when working with azetidine above ~11 eV photon energy. Literature data Anicich (1993) indicate that the ethylene radical cation C$_2$H$_4^{\bullet+}$ reacts with C$_2$H$_2$ with an overall rate $k = 8.4 \times 10^{-10}$ cm$^3 \cdot$molec$^{-1} \cdot$s$^{-1}$ to give the following products:

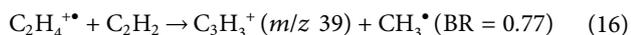
(16)

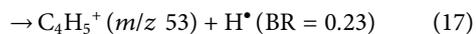
(17)

The reaction of $^{13}$CCH$_4^{+\bullet}$ should therefore lead to products at m/z 39 and 40 (i.e. C$_3$H$_3^+$ and $^{13}$CC$_2$H$_3^+$ from **Eq. 16**) plus a minor amount of m/z 54 ($^{13}$CC$_3$H$_5^+$) via **Eq. 17**.

To further check for the reactivity of $^{13}$CCH$_4^{+\bullet}$ contaminant, control experiments have been performed by selecting m/z 28 ions (a combination of HCNH$^+$ and C$_2$H$_4^{\bullet+}$) from dissociative ionization of azetidine and reacting them with C$_2$H$_2$. Products observed at m/z 39 and 53 are indicative of the occurrence of **Eqs 16, 17** and the BR is consistent with the literature value reported above. Hence contamination due to $^{13}$CCH$_4^{+\bullet}$ might overlap with products at m/z 39 and 54 in the reaction of H$_2$CNH$^{+\bullet}$ generated from azetidine. As already discussed in Sundelin et al. (2021), since the ratio of the photodissociation yields for m/z 28 and 29 at the explored photon energies is in the range 0.3–1.0, only a small contamination of $^{13}$CCH$_4^{+\bullet}$ should be present in the reagent beam.

## 4.4 Experimental Results: Data as a Function of the Photon Energy (E$_{phot}$)

Product cross sections (CSs) as a function of $E_{phot}$ for the reactions of HCNH$_2^{+\bullet}$ and H$_2$CNH$^{+\bullet}$ isomers generated through dissociative photoionization of cyclopropylamine and azetidine are given in **Figure 2**, while in **Figure 3** analogous data are presented for H$_2$CNH$^{+\bullet}$ generated via photoionization of methanimine. In both figures the large uncertainties at the lowest photon energies are due to the low parent ion intensity near threshold Richardson et al. (2021). Trends as a function of $E_{phot}$ give qualitative indications on how the reactivity is affected by changes in the internal energy content of the reagent ions.

The reaction of HCNH$_2^{+\bullet}$ (top part of **Figure 2**) gives a major product at m/z 54 with no significant dependence on $E_{phot}$ and, by extension, on the internal energy of the ion, at least up to ~12.5 eV. The decrease in CS at $E_{phot} > 12.5$ eV is an artefact due to the increasing contamination of C$_2$H$_5^+$ in the parent beam (as discussed in **Section 4.3**), which reduces the relative amount of HCNH$_2^{+\bullet}$ ion available for reaction. The other products at m/z 28 and 39 are minor, but they also do not show any significant change as a function of the photon energy, with the rise in the m/z 39 product above $E_{phot} > 12.5$ being due to the emergence of the C$_2$H$_5^+$ contaminant.

The reaction with H$_2$CNH$^{+\bullet}$ from azetidine (bottom part of **Figure 2**) also gives a major product at m/z 54, but differently from the previous isomer, its CS shows a marked decrease with increasing $E_{phot}$ before plateauing, thus pointing to a barrierless pathway somehow inhibited by the increase in internal energy of the reactant ion. The other main channel is at m/z 28, with the CS showing no change with increasing $E_{phot}$, making it the major product at $E_{phot}$ above ~11.7 eV. The m/z 39 channel shows a very small cross section up to $E_{phot} = 11.4$–11.5 eV, though above this energy a slight increase is observed which might be the result of the minor $^{13}$CCH$_4^{+\bullet}$ impurity (see discussion in **Section 4.3**). Being very minor, this channel is discussed here only due to its relevance at higher collision energies. The reaction with H$_2$CNH$^{+\bullet}$ shows an additional minor product at m/z 27, whose CS increases with increasing photon energy.

CSs for the reaction of H$_2$CNH$^{+\bullet}$ from methanimine (**Figure 3**) show the same four products, with channels at m/z 54 and 28 being the major ones at all photon energies, and those at m/z 39 and 27 being minor throughout the whole photon energy range, in good agreement with the data collected using azetidine. As already mentioned, products observed using methanimine as precursor do not suffer from the contamination issues mentioned in **Section 4.3**, however a quantitative comparison as a function of $E_{phot}$ should be performed with care, since the photon energy scales are not directly related. In fact, while in the case of methanimine photoionization the excess energy carried by the photon will





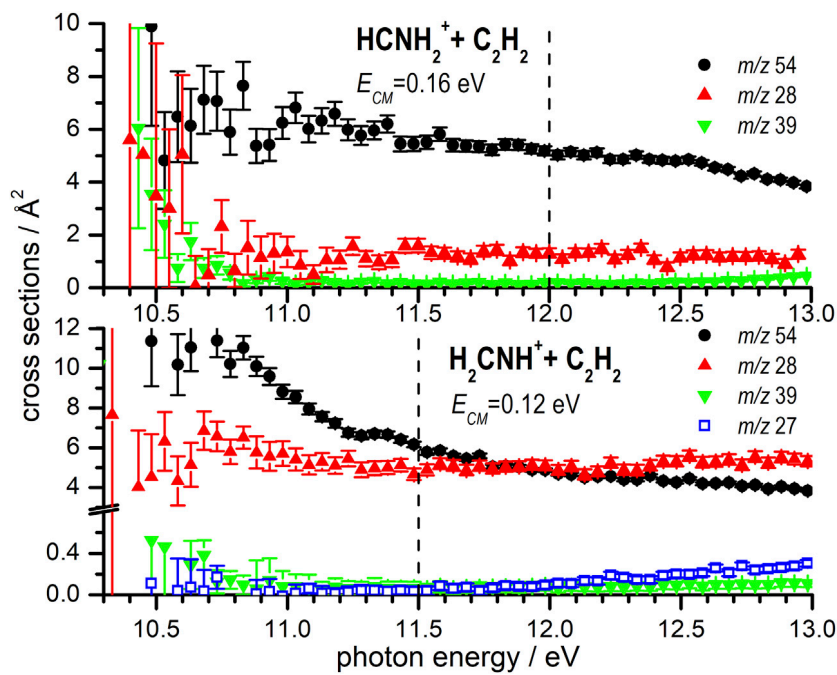

**FIGURE 2** | Reactive cross sections as a function of $E_{phot}$ for the reaction of HCNH$_2$$^{+\bullet}$ **(top)** and H$_2$CNH$^{+\bullet}$ **(bottom)** with C$_2$H$_2$. The experimental AEs of parent ions at m/z 29 from both precursors are at $E_{phot}$ = 10.2 ± 0.1 eV (outside the x axis range). Collision energies are fixed at $E_{CM}$ = 0.16 eV **(top)** and 0.12 eV **(bottom)**. The vertical dashed lines indicate the photon energies at which data have been collected as a function of $E_{CM}$ (see **Figure 4**).

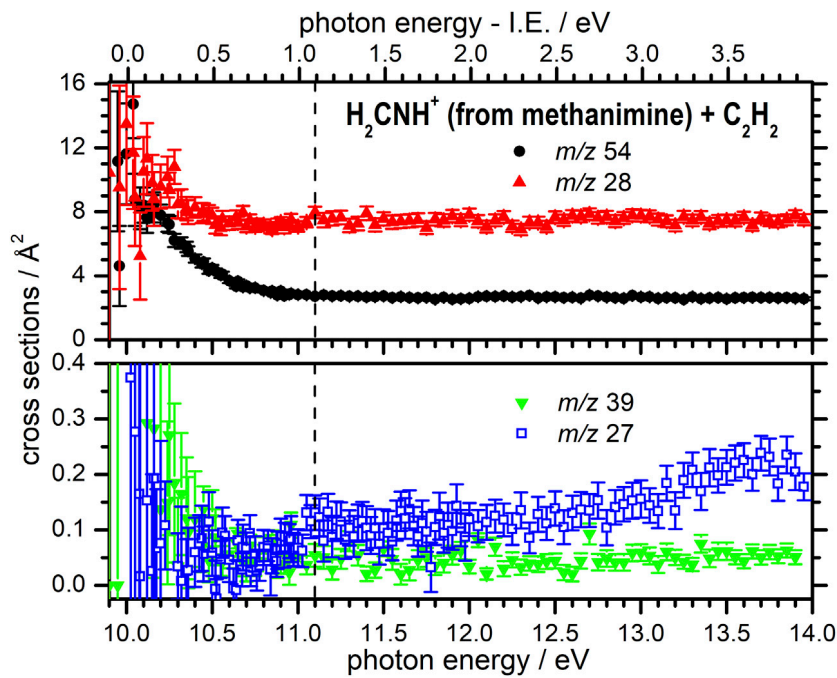

**FIGURE 3** | Reactive cross sections as a function of $E_{phot}$ for the reaction of H$_2$CNH$^{+\bullet}$ with C$_2$H$_2$. The collision energy is fixed at $E_{CM}$ ~ 0.1 eV. The x axis scale at the top is the difference between $E_{phot}$ (bottom x scale) and the AE for production of the parent ion from photoionization of methanimine, and equal to 10.01 ± 0.08 eV, as detailed in the text. The vertical dashed lines indicate the photon energies at which data have been collected as a function of $E_{CM}$ (see **Figure 5**).

**TABLE 2** | Total rate constants at fixed average energy $k_{tot}$ ($E_{ave}$) and branching ratios (BRs) for the reaction of HCNH$_2$$^{+\bullet}$ and H$_2$CNH$^{+\bullet}$ with C$_2$H$_2$. Results have been obtained at $E_{CM}$ = 0.08 ± 0.01 eV, corresponding to an average energy $E_{ave}$ = 0.10 ± 0.01. The HCNH$_2$$^{+\bullet}$ isomer was produced via dissociative photoionization of cyclopropylamine at $E_{phot}$ = 12.0 eV while the H$_2$CNH$^{+\bullet}$ isomer was produced by photoionization of methanimine at $E_{phot}$ = 11.1 eV.

|  | HCNH$_2$$^{+\bullet}$ | H$_2$CNH$^{+\bullet}$ |
|---|---|---|
| $k_{tot}$ ($E_{ave}$)[a] | (1.0 ± 0.2) × 10$^{-10}$ | (1.2 ± 0.2) × 10$^{-10}$ |
| ($k_{tot}$/$k_L$)[b] | 0.09 | 0.1 |
| product (m/z) | Branching ratios (BRs) | |
| 28 | 0.22 ± 0.06 | 0.74 ± 0.15 |
| 54 | 0.74 ± 0.16 | 0.25 ± 0.05 |
| 27 | n.d[c] | 0.010 ± 0.003 |
| 39 | 0.03 ± 0.01 | <0.005 |

[a]Total rate constant (in cm$^3$·molec$^{-1}$·s$^{-1}$) at the specified average collision energy $E_{ave}$, estimated as detailed in the text.
[b]For both isomers $k_L$ = 1.18 × 10$^{-9}$cm$^3$·molec$^{-1}$·s$^{-1}$ is the Langevin rate constant, calculated using the value 3.487 Å$^3$ for the average electronic polarizability of C$_2$H$_2$.
[c]Non detected.

contribute to increase the internal energy of the reacting H$_2$CNH$^{+\bullet}$, in the case of dissociative photoionization from azetidine the excess energy might be dissipated also as internal and kinetic energy of the neutral co-fragment. Therefore, the change in internal energy of H$_2$CNH$^{+\bullet}$ with increasing $E_{phot}$ is expected to be smaller in the azetidine experiment than for the methanimine precursor. CS for product at m/z 54 decreases sharply with increasing $E_{phot}$ up to about 10.8 eV, after which it levels off and is approximately constant. Possible explanations of the observed trend are discussed in **Section 6**, here we only highlight the good qualitative agreement with the decrease as a function of $E_{phot}$ observed with the dissociative photoionization of azetidine.

Similarly to the azetidine experiment, the CS for the m/z 28 product shows no notable change as a function of $E_{phot}$. The CS for the very minor product at m/z 39 also shows no significant dependence on $E_{phot}$. This is at odds with the azetidine data in which a very slight increase was observed at higher photon energies, further supporting the idea that this increase is due to the emergence of a pathway with the $^{13}$CCH$_4$$^{+\bullet}$ impurity. Finally, CS for the other minor channel at m/z 27 shows a similar rise with increasing $E_{phot}$ as observed with the azetidine precursor.

The BRs for the various product channels, obtained from the absolute cross section measurements, are summarized in **Table 2** for both isomers. In the same Table estimates of the total (i.e. summed over all the product channels) rate constants at a fixed average energy $k_{tot}$ ($E_{ave}$) have been given. For details on how cross sections have been converted into rate constants see our previous paper Sundelin et al. (2021). For the HCNH$_2$$^{+\bullet}$ isomer, BRs and $k_{tot}$ values were obtained via dissociative photoionization of cyclopropylamine at $E_{phot}$ = 12.0 eV, while for the H$_2$CNH$^{+\bullet}$ the methanimine data at $E_{phot}$ = 11.1 eV were used. For both isomers the collision energy was fixed at $E_{CM}$ = 0.08 ± 0.01 eV, corresponding to an average energy $E_{ave}$ = 0.10 ± 0.01 eV Ervin and Armentrout (1985), Nicolas et al. (2002). The ratio between $k_{tot}$ and the Langevin collision rate constant ($k_L$), also reported in **Table 2**, is a useful estimate of the overall efficiency of the reactions, which proceed at a rate of ~10% of the Langevin limit for both isomers.

## 4.5 Experimental Results: Data as a Function of the Collision Energy ($E_{CM}$)

The CSs as a function of $E_{CM}$ for the reaction of both HCNH$_2$$^{+\bullet}$ and H$_2$CNH$^{+\bullet}$ isomers generated through dissociative photoionization of cyclopropylamine and azetidine are shown in **Figure 4**, while in **Figure 5** analogous data are presented for H$_2$CNH$^{+\bullet}$ generated via photoionization of methanimine.

For the HCNH$_2$$^{+\bullet}$ isomer, the m/z 54 product shows a marked decrease with increasing $E_{CM}$, indicative of a barrierless pathway most likely proceeding via a complex-mediated mechanism. The m/z 28 and 39 channels show a combination of two trends, with a similar decrease below 1 eV followed by a gradual increase with increasing $E_{CM}$, indicating the presence of a barrierless pathway, mostly contributing at low collision energy, superimposed on an endothermic process (or a process involving a barrier) appearing at high $E_{CM}$. It should be noted that CS at m/z 28 are affected by collisional fragmentation of the parent which is discussed in more detail below.

For the H$_2$CNH$^{+\bullet}$ isomer with the azetidine precursor (bottom of **Figure 4**), CSs were measured only for products at m/z 54 and 39, with the former decreasing with increasing $E_{CM}$ and the latter showing a decrease with increasing $E_{CM}$, at low collision energies, followed by a ramp at very high collision energies. A more complete picture for the reactivity of the H$_2$CNH$^{+\bullet}$ isomer is obtained from the experiments using methanimine (see **Figure 5**) since CS as a function of $E_{CM}$ have been measured for all of the four product channels. The trend for m/z 54 product, decreasing with increasing $E_{CM}$, confirms the results from the azetidine experiment. Also similar to the azetidine experiment, the m/z 39 product is minor at all collision energies and shows an initial decrease with increasing $E_{CM}$ followed by an inversion in the $E_{CM}$ trend at high energies. The m/z 28 channel shows a slight initial decrease followed by a gradual increase with increasing $E_{CM}$ that resembles the trend observed for the HCNH$_2$$^{+\bullet}$ isomer. The m/z 27 product shows a similar increase with increasing $E_{CM}$ as it does with photon energy, indicating that the barrier to this pathway can be overcome equally effectively with collision energy as internal energy. A detailed discussion with explanations of the trends as a function of collision energy is given in **Section 6** on the basis of the results obtained from computations and presented in the next section.

## 5 COMPUTATIONAL RESULTS

Our calculations show that HCNH$_2$$^{+\bullet}$ is slightly more stable (by 18.9 kJ/mol) than H$_2$CNH$^{+\bullet}$, consistent with a previous literature value Nguyen et al. (1994) in which the difference in the heat of formations of the two isomers, calculated at the UQCISD(T)/6-311++G(3df, 2p) and RCCSD(T)/cc-pVTZ levels of theory, is





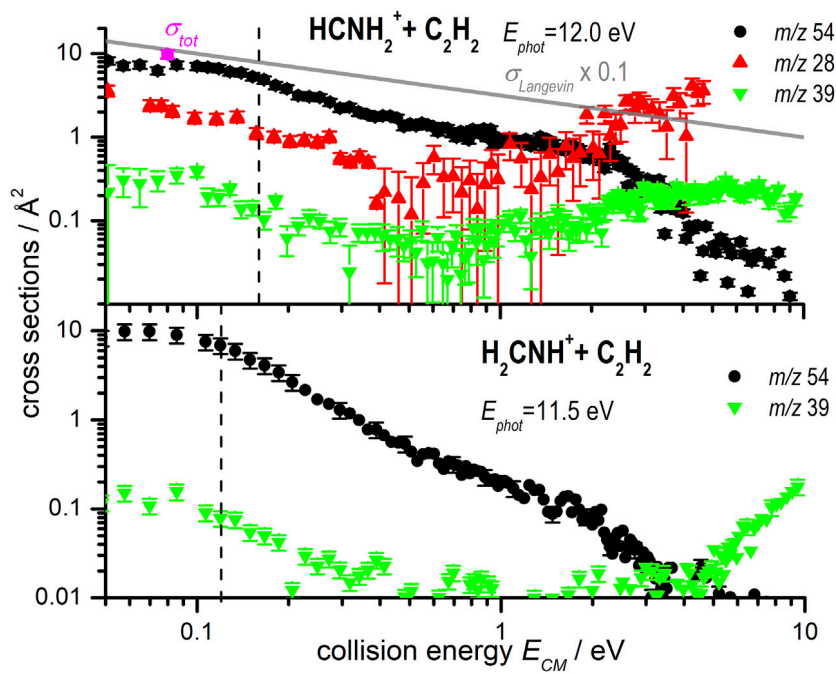

**FIGURE 4** | Reactive cross sections as a function of the collision energy ($E_{CM}$) for the reaction of $HCNH_2^{+\bullet}$ **(top)** and $H_2CNH^{+\bullet}$ **(bottom)** with $C_2H_2$. The photon energy is fixed at $E_{phot}$ = 12.0 eV **(top)** and 11.5 eV **(bottom)**. The vertical dashed lines indicate the collision energy at which data have been collected as a function of $E_{phot}$ (see **Figure 2**). In the top panel: the grey solid line is the Langevin CS (rescaled by a factor 0.1 to fit in the figure). The magenta point is the total reactive CS ($\sigma_{tot}$) used to calculate the total rate constant $k_{tot}$ reported in **Table 2**.

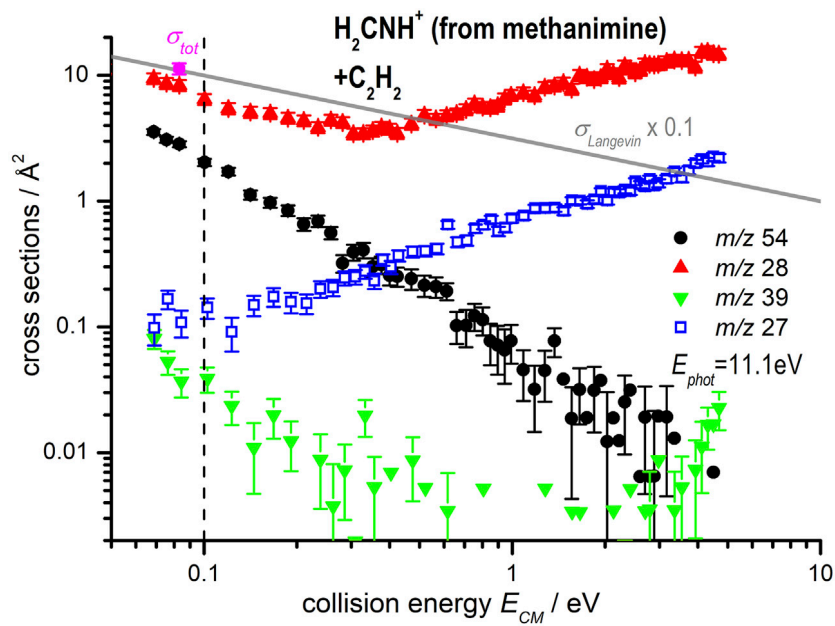

**FIGURE 5** | Reactive cross sections as a function of $E_{CM}$ for the reaction of $H_2CNH^{+\bullet}$ (from methanimine) with $C_2H_2$. The photon energy is fixed at $E_{phot}$ = 11.1 eV. The vertical dashed line is at the collision energy at which data have been collected as a function of $E_{phot}$ (see **Figure 3**). The grey solid line is the Langevin CSs (rescaled by a factor 0.1 to fit in the figure). The magenta point is the total reactive CS ($\sigma_{tot}$) used to calculate the total rate constant $k_{tot}$ reported in **Table 2**.





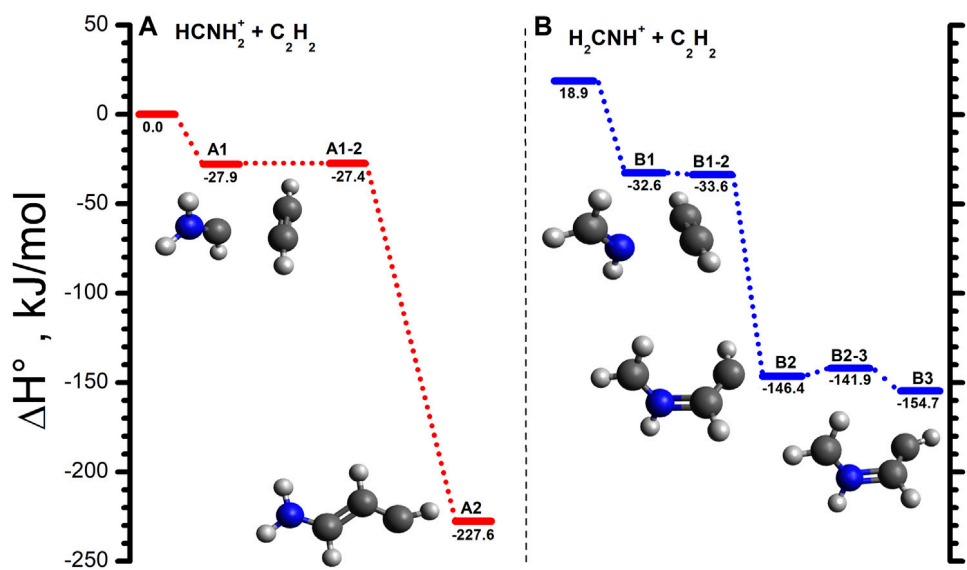

**FIGURE 6 |** Structures and energies of the van der Waals adducts (and their evolution into covalently bound intermediates) for the entrance channel of the reactions of HCNH$_2$$^{+\bullet}$ (panel A) and H$_2$CNH$^{+\bullet}$ (panel B) plus C$_2$H$_2$. Energies (in kJ/mol) are defined with respect to the separated HCNH$_2$$^{+\bullet}$ plus C$_2$H$_2$ reactants. Calculations are at the MP2/6-311++G(d,p)//CCSD(T)/6-311++G(d,p) level of theory.

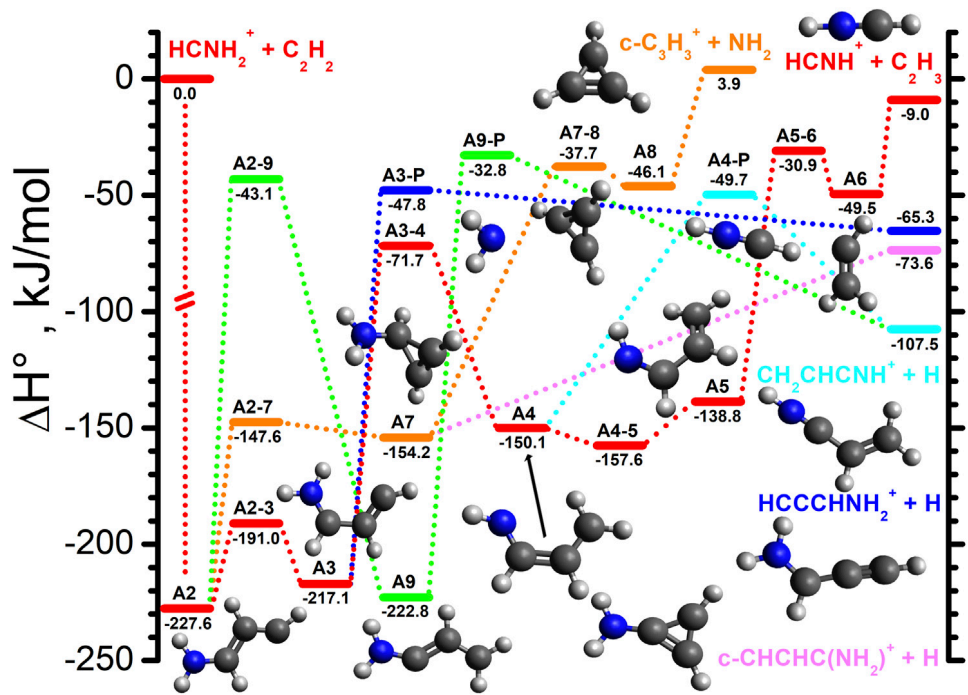

**FIGURE 7 |** Energies and reaction pathways for the HCNH$_2$$^{+\bullet}$ plus C$_2$H$_2$. Energies (in kJ/mol) are defined with respect to the separated HCNH$_2$$^{+\bullet}$ plus C$_2$H$_2$ reactants. Calculations are at the MP2/6-311++G (d,p)//CCSD(T)/6-311++G(d,p) level of theory. Molecular structures are shown only for intermediates and products. For TSs' structures see the **Supplementary Material**.

equal to 16 kJ/mol. Given the open shell nature and the presence of a charge such differences in the calculations are not surprising. All enthalpies are calculated with the GAUSSIAN suite of programs at the CCSD(T)/6-311++G(d,p) level of theory.

Unless explicitly noted otherwise, relative enthalpies are given (in kJ/mol and in parenthesis after each structure) with respect to the sum of the separated reactants having the lowest energy, i.e. HCNH$_2$$^{+\bullet}$ plus C$_2$H$_2$.

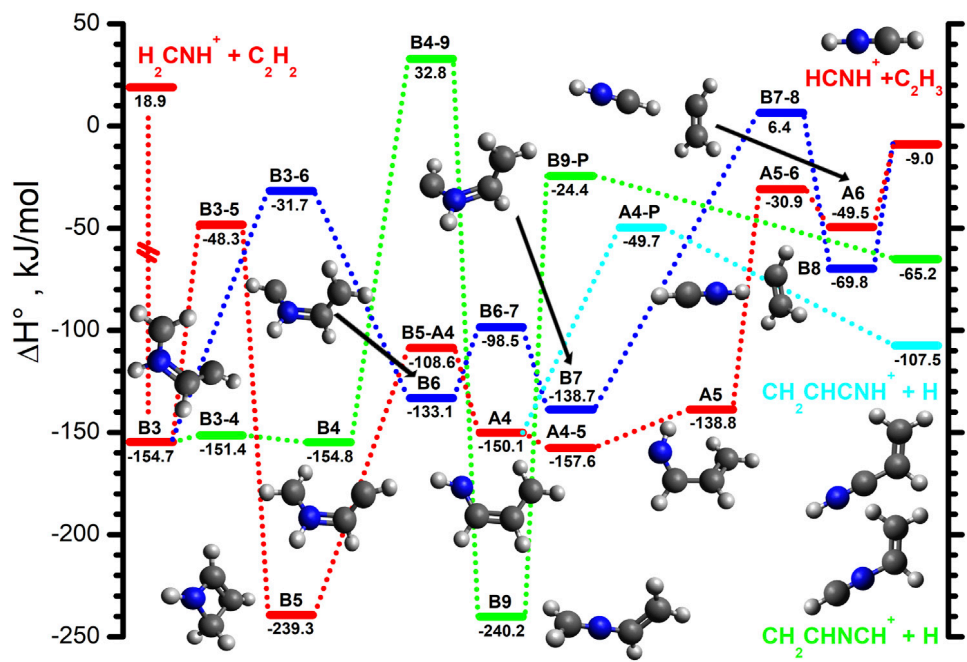

**FIGURE 8** | Energies and reaction pathways for the H₂CNH⁺• plus C₂H₂. Energies (in kJ/mol) are defined with respect to the separated HCNH₂⁺• plus C₂H₂ reactants. Calculations are at the MP2/6-311++G(d,p)//CCSD(T)/6-311++G(d,p) level of theory. Molecular structures are shown only for intermediates and products. For TSs' structures see the **Supplementary Material**.

For both isomers, the reaction mechanisms proceed via the formation of covalently bound adducts through a radical-terminal approach, with the barriers for the non-radical terminal attack being endothermic by more than 45 kJ/mol in both cases.

**HCNH₂⁺• isomer** (pathways are described graphically in **Figures 6**, **7**, while reaction enthalpies are compared with literature values in **Table 1**): for this isomer, an initial van der Waals cluster **A1** (−27.9) can proceed to a covalently-bound adduct **A2** (−227.6) via the transition state (TS) **A1-2** (−27.4). From **A2**, a series of pathways are possible, all of which are shown diagrammatically in **Figure 7** as well as being described here.

The most favorable pathway, from the thermochemical point of view, to form the observed major product at m/z 54 proceeds from a rotamer of **A2**, named **A3** (−217.1), with the two structures being separated by the TS **A2-3** (−191.0). **A3** can then undergo a [1,4] H-shift (from the N atom to the terminal C) to give **A4** (−150.1) via the TS **A3-4** (−71.7). From **A4** the ejection of an H• via **A4-P** (−49.7) gives the m/z 54 product in the form of the lowest energy isomer H₂CCHCNH⁺ (protonated vinyl cyanide) in combination with H• (−107.5).

While producing a higher energy C₃H₄N⁺ isomer, the lowest energy pathway for m/z 54 proceeds through the cyclic intermediate **A7** (−154.2) which can be reached via **A2-7** (−147.6). **A7** can then barrierlessly eject the H• from the central carbon to give a three-membered ring aminic ion c-CHCHC(NH₂)⁺ (−73.6).

A different C₃H₄N⁺ isomer at m/z 54 can be formed from **A3** by the direct ejection of H• via the TS **A3-P** (−47.8) to give a linear aminic ion HCCCHNH₂⁺ plus H• (−65.3). A fourth pathway leading to the formation of a product at m/z 54 proceeds from **A2**, which can undergo a [1,3] H-shift via the TS **A2-9** (−43.1) to give **A9** (−222.8), which can then eject H• via TS **A9-P** (−32.8) to give the already mentioned H₂CCHCNH⁺ plus H•. All these pathways are fully submerged, hence they can all contribute to the observed m/z 54 product.

Though not included in **Figure 7**, our calculations have identified another pathway, higher in energy with respect to the ones above, that leads to the production of a fourth high energy C₃H₄N⁺ isomer, the linear CH₂CCNH₂⁺ (protonated ketenimine). This pathway (shown in **Supplementary Figure S30** of the Supplementary Material) starts from **A9** and proceeds via successive interconversions into the rotamers **A10** (−230.5), via TS **A9-10** (−226.7), and then into **A11** (−180.1), via TS **A10-11** (−188.9), with the lower energy of the TS being due to the zero-point correction. **A11** can undergo a hydrogen migration to give **A12** (−226.2), via TS **A11-12** (−60.5), which can finally eject H• to give CH₂CCNH₂⁺ plus H• (−23.6) via TS **A12-P** (−3.3).

The next major product from the HCNH₂⁺• isomer is at m/z 28. A possible pathway has been identified stemming from **A4**. This intermediate can interconvert to give its rotamer **A5** (−138.8) via the TS **A4-5** (−157.6). Note that the lower energy of the TS compared to **A4** is due to the zero point correction. **A5** can then cleave the central C-C bond to give a van der Waals complex of the products, **A6** (−49.5), via the TS **A5-6** (−30.9). The van der Waals adduct can subsequently fragment, with no energy barrier, into the products HCNH⁺ (protonated hydrogen cyanide) plus C₂H₃• (−9.0).

Finally, a pathway should be accounted for the extremely minor product detected at m/z 39 and attributed to c-C₃H₃⁺. It starts from the cyclic **A7** intermediate, which can cleave the C-N bond to give the van der Waals complex **A8** (−46.1) via the TS **A7-8** (−37.7). **A8** can then fragment barrierlessly into the products c-C$_3$H$_3$$^+$ and NH$_2$$^{+\bullet}$ (+3.9). The small yield of the *m/z* 39 product is compatible with the calculated endothermicity of the channel.

**H$_2$CNH$^{+\bullet}$ isomer** (pathways are described graphically in **Figures 6**, **8**, while reaction enthalpies are compared with literature values in **Table 1**): for this isomer the reagents (at +18.9 kJ/mol above the other isomer) come together to form an initial van der Waals complex **B1** (−32.6) which can then evolve, by formation of a C-N bond, into the covalently-bound complex **B2** (−146.4) via the TS **B1-2** (−33.6), with the lower energy of the TS compared to the van der Waals complex arising from the zero-point energy correction. **B2** can then interconvert into its rotamer **B3** (−154.7) via TS **B2-3** (−141.9) from which the various pathways then proceed, as shown diagrammatically in **Figure 8** and described in the following.

The lowest energy pathway to form the *m/z* 54 product proceeds via a cyclisation to give a four-membered ring intermediate, **B5** (−239.3), via **B3-5** (−48.3). **B5** can open to give the N-terminal structure **A4**, via **B5-A4** (−108.6). The **A4** intermediate is a common structure with the potential energy hypersurface of the other isomer, hence from there the reaction can proceed to give H$_2$CCHCNH$^+$ (protonated vinyl cyanide) plus H$^{\bullet}$ (−107.5), as already described above. Other C$_3$H$_4$N$^+$ isomers could be produced from **B3** via an interconversion into a second rotamer **B4** (−154.8) via the TS **B3-4** (−151.4). The latter rotamer can undergo a [1,3] H-shift to give **B9** (−240.2) via **B4-9** (+32.8). **B9** can then eject H$^{\bullet}$ to give H$_2$CCHNCH$^+$ (protonated vinyl isonitrile) plus H$^{\bullet}$ (−65.2) via the TS **B9-P** (−24.4). While this pathway involves a barrier higher in energy than the reactants (*i.e.* **B4-9** that is 13.9 kJ/mol above the reactants) it is potentially relevant at higher photon energies where the HCNH$_2$$^{+\bullet}$ isomer might be formed with sufficient internal energy to both overcome the barrier and to potentially inhibit the alternative low energy pathway going via the cyclisation to give **B5**.

There are also two higher-energy pathways leading to another C$_3$H$_4$N$^+$ isomer at *m/z* 54 that proceed via **B9**. For clarity reasons they have not been included in **Figure 8**, but they are shown in **Supplementary Figure S31** and are described briefly here. The first pathway involves a [1,2] H-shift to give **B11** (−277.1) via a TS **B9-11** (−35.5), with **B11** then ejecting H$^{\bullet}$ to give CH$_2$CNCH$_2$$^+$ (−24.4) via **B11-P** (+3.7). The second involves a [1,3] H-shift from **B9** to give **B12** (-276.1) via TS **B9-12** (+10.2), which then also eject H$^{\bullet}$ to give CH$_2$CNCH$_2$$^+$, this time via TS **B12-P** (+3.2).

The *m/z* 28 product, corresponding to HCNH$^+$ formed in association with C$_2$H$_3$$^{\bullet}$, can proceed from the **A4** intermediate, as already described for the HCNH$_2$$^{+\bullet}$ isomer, but it can also form from **B3** via a [1,4] H-shift to give **B6** (−133.1) via the TS **B3-6** (−31.7). **B6** can then interconvert into its rotamer **B7** (−138.7), via the TS **B6-7** (−98.5), which can then cleave the central C-N bond to give **B8** (−69.8), a van der Waals complex of the products, via **B7-8** (6.4). **B8** can then dissociate barrierlessly into HCNH$^+$ and C$_2$H$_3$$^{\bullet}$.

As **A4** is a common structure, the whole of the HCNH$_2$$^{+\bullet}$ PES is also accessible for the H$_2$CNH$^{+\bullet}$ isomer, therefore creating a potential pathway to a *m/z* 39 product through a series of rearrangements via **A2**, as previously described for the HCNH$_2$$^{+\bullet}$ isomer. However, due to the large number of interconversions, any contribution from this pathway is expected to be extremely minor despite being fully submerged with respect to the reactants.

While not included in **Figure 8**, there is also a pathway to interconvert between **B5** and the rotamers **B6** and **B7** and a diagram is shown in **Supplementary Figure S32**. **B5** can undergo a [1,2] H-shift to give **B10** (−231.4) via **B5-10** (−71.7). This can then ring-open via C-C bond cleavage to give **B6** via **B10-6** (−48.3) or **B7** via **B10-7** (−24.3). For the very minor *m/z* 27 product, only observed at high photon and high collision energies, two endothermic pathways have been found, one proceeding via intermediate **B6** without a barrier to give C$_2$H$_3$$^+$ plus HCNH (+145.1) and the other one going from **B9** to give C$_2$H$_3$$^+$ plus H$_2$CN (+104.0), also with no further barrier. Since the energies of both products are out of scale they have not been reported in **Figure 8**.

## 6 DISCUSSION

Comparison of the two sources of H$_2$CNH$^{+\bullet}$ shows a high level of qualitative agreement with regards to the trends as a function of both $E_{phot}$ and $E_{CM}$, as well as broad quantitative agreement for both CSs and BRs of the various products. Exact quantitative comparison is not possible due to the different scaling of internal energy with photon energy. Similarly, comparison of the data collected with the two different isomers indicates a high level of selectivity for both isomers over the whole photon energy range.

For both isomers the major product is at *m/z* 54, corresponding to C$_3$H$_4$N$^+$. Previous computational studies Wang et al. (2015), Heerma et al. (1986) have provided strong evidence for the existence of several C$_3$H$_4$N$^+$ isomers in the gas phase, the most stable of which being protonated vinyl cyanide H$_2$CCHCNH$^+$. The next most stable is a three-membered ring structure c-(CHCHC)–NH$_2$$^+$ that is 22.2 kJ/mol higher in energy Wang et al. (2015), followed by the linear aminic structure HCCCHNH$_2$$^+$ [37.7 kJ/mol above the most stable isomer Wang et al. (2015)]. At higher energies there is protonated vinyl isonitrile, H$_2$CCHNCH$^+$ [49.8 kJ/mol above the most stable isomer Wang et al. (2015)], protonated ketenimine H$_2$CCCNH$_2$$^+$ [57.4 kJ/mol above the most stable isomer Wang et al. (2015)] and CH$_2$CNCH$_2$$^+$ [61.1 kJ/mol above the most stable isomer Wang et al. (2015)]. Although our experimental technique can not distinguish between the different C$_3$H$_4$N$^+$ isomers, *ab-initio* calculations can help in differentiating among the pathways leading to them.

For the reaction of HCNH$_2$$^{+\bullet}$ the *m/z* 54 product is approximately constant in cross section as a function of the photon energy while it decreases with increasing collision energy. These findings can be rationalized by the fact that the *m/z* 54 formation pathways have only submerged barriers and proceed via complex-mediated mechanisms, as confirmed by our calculations. In fact, not just one, but four different pathways with fully submerged barriers have been found (see **Section 5**) to yield the three lowest energy isomers $CH_2CHCNH^+$, c-$CHCHC(NH_2)^+$ and $HCCCHNH_2^+$. Hence, the observed m/z 54 product is expected to correspond to a combination of the three isomers.

For the $H_2CNH^{+\bullet}$ isomer, the CS trend with $E_{phot}$ for the m/z 54 product can be explained through a combination of multiple channels. At low internal energies, a barrierless pathway proceeds via the cyclized adduct **B5** but this decreases in relevance as the internal energy increases due to the inhibition of the cyclisation. At higher $E_{phot}$, a pathway with a slight barrier via **B9** opens, with this pathway giving the plateau that is observed at higher internal energies. Hence, for $H_2CNH^{+\bullet}$, in the absence of internal excitation, the observed m/z 54 product is exclusively due to protonated vinyl cyanide $H_2CCHCNH^+$.

As a function of the collision energy, the m/z 54 channels for both isomers show an initial decrease in CS section, indicative of barrierless pathways as suggested by our calculations. This is due to the higher collision energy inhibiting the formation of the van der Waals clusters (**A1** and **B1**) in the entrance channels.

For the reaction of $HCNH_2^{+\bullet}$, the m/z 28 and 39 products are approximately constant in cross section as a function of $E_{phot}$. This is in agreement with the computations that identify fully submerged pathways for the m/z 28 product and for formation of the van der Waals complex of the products for the m/z 39 channel, with the barrierless fragmentation of this complex lying approximately equal in energy with the reactants. The lower cross section of the m/z 28 product for this isomer than for $H_2CNH^{+\bullet}$ is expected to be due to the presence of the multiple lower-lying channels to give products at m/z 54 which are discussed above.

For the reaction of $H_2CNH^{+\bullet}$, the m/z 28 channel corresponds to a combination of collision induced dissociation, especially at higher collision energies, with the two fully submerged pathways (see blue and red lines in **Figure 8**) leading to a CS trend that is independent of the photon energy at low collision energies. It is not clear what contribution there might be at low collision energies from the fragmentation of the parent at higher photon energies.

The small signal observed for the m/z 39 product for $H_2CNH^{+\bullet}$ is expected to indicate the very small level of interconversion between the isomers via **A4**. While this interconversion is barrierless for both ions, the fact that this requires passage over twelve TSs and intermediate structures is expected to lead to a significant level of kinetic inhibition, especially compared to other barrierless pathways present.

In general, products at m/z 28 and 39 from both isomers show an initial decrease in CS section with increasing collision energy, indicative of barrierless pathways as suggested by our calculations. This is due to the higher collision energy inhibiting the formation of the van der Waals clusters (**A1** and **B1**) in the entrance channels. For both isomers, above ~0.7 eV, the m/z 28 product shows a gradual rise with increasing collision energy which is probably due to the collisional fragmentation of the parent ion into $HCNH^+$ and $H^\bullet$. The very minor m/z 39 channel also shows a higher collision energy feature for both isomers, with appearance energies of ~1 eV for $HCNH_2^{+\bullet}$ and ~4 eV for $H_2CNH^{+\bullet}$. This is believed to correspond to a mechanism to give either c-$C_3H_3^+$ or its high energy isomer $CH_2CCH^+$ via a significant barrier, though no individual channel has been identified.

The m/z 27 channel is only observed with the $H_2CNH^{+\bullet}$ isomer, where it shows an increase with increasing photon energy, a trend that relates well with the endothermic channels **Eqs 3** and **4** identified both in literature and our calculations. As a function of the collision energy, this channel shows a steady increase, again indicative of an endothermic proton transfer process which can proceed directly at higher collision energies.

## 7 CONCLUSION

Absolute cross sections for the reaction of methanimine ($H_2CNH^{+\bullet}$) and aminomethylene ($HCNH_2^{+\bullet}$) radical cations with $C_2H_2$ have been measured as a function of photon and collision energies under a single collision regime using a guided ion beam tandem mass spectrometer with VUV ionization for ion generation.

Through comparison of $H_2CNH^{+\bullet}$ formed both via direct and dissociative ionization and $HCNH_2^{+\bullet}$ formed exclusively through dissociative ionization, we are able to conclude that both isomers are formed selectively with a high level of correspondence between the two $H_2CNH^{+\bullet}$ sources. *Ab-initio* calculations of the most relevant stationary points on the potential energy surfaces allow for the rationalization of experimental results.

Our experiments clearly show that the reaction of both isomers with $C_2H_2$ lead to the formation of $C_3H_4N^+$ products (detected at m/z 54), and by extension to an elongation of the main skeleton of the ion. Measured CSs show a negative dependence on the collision energy, indicative of barrierless and exothermic processes that are feasible at low temperatures. Thus, the title reactions can be viable intermediate steps to produce larger ions in the ISM as well as ionospheres of planets and their satellites.

But what about the structures of the $C_3H_4N^+$ products? Theoretical calculations show that, for the reaction of $HCNH_2^{+\bullet}$, the lowest energy isomer ($H_2CCHCNH^+$, protonated vinyl cyanide, a.k.a. acrylonitrile) as well as two higher energy isomers, namely c-$CHCHC(NH_2)^+$ and $HCCCHNH_2^+$, can be generated via barrierless and exothermic pathways. It should, however, be noted that our experiment cannot yield any information about the branching ratios for the formation of the different $C_3H_4N^+$ isomers. Since all of these pathways involve submerged barriers, no preference for the production of individual ionic products could be inferred from our calculations starting from the $HCNH_2^{+\bullet}$ isomer. Conversely, for the $H_2CNH^{+\bullet}$ reagent, the only barrierless pathway is the one leading to the production of protonated vinyl cyanide, which somewhat restricts the potential of the methanimine radical cation in the synthesis of interstellar nitrogen compounds.

Protonated vinyl cyanide has been proposed as a likely intermediate in star forming regions and in the atmosphere of Titan, where the title reaction can contribute to its formation. In particular, Cassini flyby missions to Titan's upper atmosphere have detected the presence of a strong signal at *m/z* 54 that has been tentatively assigned to protonated vinyl cyanide on the basis of the large proton affinity of the neutral Vuitton et al. (2007). More recently, the spectroscopic detection of vinyl cyanide in Titan's atmosphere has been confirmed using ALMA data Palmer et al. (2017). In the ISM, while vinyl cyanide is one of the most abundant nitriles detected in several environments such as the circumstellar shell of evolved carbon stars and dark clouds, the presence of its protonated form has been proposed on the basis of laboratory spectra but still not confirmed Martinez et al. (2013). It is therefore noteworthy that our title reaction, starting from the methanimine radical cation (H$_2$CNH$^{+\bullet}$), could lead to protonated vinyl cyanide, from which acrylonitrile can be formed either through dissociative recombination with electrons Vigren et al. (2009) or proton transfer to species such as NH$_3$ Taquet et al. (2016).

## DATA AVAILABILITY STATEMENT

The original contributions presented in the study are included in the article/**Supplementary Material**. Datasets shown in Figures 2-5 are made available as ASCII files in the Zenodo repository accessible via DOI: 10.5281/zenodo.5419565. Further inquiries can be directed to the corresponding author.

## AUTHOR CONTRIBUTIONS

J-CG synthesized the reagent methanimine. CA, CR, RT, and JZ developed the experimental set-up and RT developed the data acquisition and analysis suite. All authors were involved in collecting the data. VR performed the data analysis of the experimental data. WG and DS worked on the PESs for both isomers while PT worked on the PESs for the fragmentation of reagent species. DA and VR worked on the interpretation of the results while DA, WG, and VR wrote the manuscript draft. All authors were involved in reviewing the manuscript.

## ACKNOWLEDGMENTS

We are grateful to the DESIRS beamline team for their assistance during the synchrotron measurements and to the technical staff of SOLEIL for the smooth running of the facility under projects nos. 20180118 and 20190249. This work was supported by the European Union's Horizon 2020 research and innovation program "Astro-Chemistry Origins" (ACO), Grant No. 811312. WG thanks the Swedish Research Council for a project grant (grant number 2019-04332). MP and JZ acknowledge support from the Ministry of Education, Youth and Sports of the Czech Republic (grant No. LTC20062). VR acknowledges funding for a PhD fellowship from the Dept. Physics, University of Trento. J-CG thanks the Centre National d'Etudes Spatiales (CNES) for a grant. CA, RT and CR acknowledge the synchrotron SOLEIL for the support to the associated CERISES setup since 2008 and subsistence expenses during beamtime periods. The Orsay group acknowledges support from the LabEx PALM (ANR-10-LABX-0039-PALM, Project ERACOP).

## SUPPLEMENTARY MATERIAL

The Supplementary Material for this article can be found online at: https://www.frontiersin.org/articles/10.3389/fspas.2021.752376/full#supplementary-material